\documentclass{iopart}
\usepackage{iopams, setstack}
\usepackage{bm,hyperref}

\newcommand{\eqref}[1]{{(\ref{#1})}}    

\newcommand{\rmSelf}{{\rm self}}    
\newcommand{\bfx}{{\mathbf{x}}} 
\newcommand{\ItP}{{\mathcal{P}}} 

\newcommand{\Gret}{{G_{\mathrm{ret}}}}  
\newcommand{\Gadv}{{G_{\mathrm{adv}}}}  
\newcommand{\GS}{{G_{\mathrm{S}}}}     
\newcommand{\GSDW}{{G_{\mathrm{S,DW}}}}     
\newcommand{\GR}{{G_{\mathrm{R}}}}      

\newcommand{\PhiRet}{{\Phi_{\mathrm{ret}}}}    
\newcommand{\PhiR}{{\Phi_{\mathrm{R}}}}  
\newcommand{\PhiS}{{\Phi_{\mathrm{S}}}}  
\newcommand{\phiS}{{\phi^\rmSelf}}  

\newcommand{\Lie}{{\mathcal{L}}} 
\newcommand{\LieX}{{\mathcal{L}_{\xi}}} 

\begin{document}
\title{Self-forces from generalized Killing fields}
\author{Abraham I. Harte}
\address{Enrico Fermi Institute}
\address{University of Chicago, Chicago, IL 60637 USA}
\ead{harte@uchicago.edu}

\date{July 7, 2008}

\begin{abstract}
A non-perturbative formalism is developed that simplifies the
understanding of self-forces and self-torques acting on extended
scalar charges in curved spacetimes. Laws of motion are locally
derived using momenta generated by a set of generalized Killing
fields. Self-interactions that may be interpreted as arising from
the details of a body's internal structure are shown to have very
simple geometric and physical interpretations. Certain modifications
to the usual definition for a center-of-mass are identified that
significantly simplify the motions of charges with strong
self-fields. A derivation is also provided for a generalized form of
the Detweiler-Whiting axiom that pointlike charges should react only
to the so-called regular component of their self-field. Standard
results are shown to be recovered for sufficiently small charge
distributions.
\end{abstract}


\vskip 2pc

\section{Introduction}

The detailed behavior of a compact body can depend on several kinds
of interactions. These might include complicated internal contact
stresses as well as the effects of long-range gravitational and
electromagnetic fields. Restricting attention to only a few
quantities like the center-of-mass acceleration often eliminates
most of the dependence on an object's internal details. Long-range
fields largely determine the ``bulk'' motion. Such effects can often
be thought of as having two components. One is essentially imposed
by the external universe, while the other arises from the body
itself. There can be some physical ambiguity in this splitting,
although there are many systems where it provides significant
simplifications.

The discussion here focuses on the self-forces and self-torques
affecting a body's net linear and angular momenta. This problem has
a very long history. One of its interesting aspects follows from the
observation that a body's own fields strongly depend on the details
of its internal structure. Despite this apparent complication, there
exists a regime where the motion remains relatively independent of
that structure. Only a small portion of a body's self-field directly
influences its bulk motion (at least if inertial effects are
excluded). This conclusion has been reached using a number of
calculations that derive approximate self-fields for extended bodies
using perturbation theory
\cite{Abraham,Schott,Crowley,OriExtend,HarteEM}. Unfortunately, such
methods are extremely tedious, specialized, and not particularly
enlightening. Similar results are much more easily obtained by
writing down axioms for the behavior of point particles
\cite{Dirac,QuinnWald,Quinn,DetWhiting,RosenthalMassField}. It is
then assumed at the outset that only a particular portion of the
self-field affects the motion. This has been an expectation rather
than a prediction of the underlying theory.

One goal of this paper is to show that appropriately sharpened
versions of these assumptions can be derived from first principles.
An ignorable component of the self-field may be identified and
removed in the full theory. It is not necessary to appeal to
perturbation theory or the mathematical inconsistencies of point
particles. This is done by considering which portions of the
self-field satisfy an appropriate analog of Newton's third law. Any
such components cannot affect the net momenta, and may be discarded.
Geometrically, this has the interpretation of considering how a
particular Green function is deformed under the action of a
generalized Poincar\'{e} group first discussed in \cite{HarteSyms}.
This is the same group used to generate the quantities referred to
as momenta in the absence of exact Killing fields.

The modern interest in self-force problems has mainly been motivated
by the problem of extreme mass ratio binaries inspiralling under the
action of gravitational radiation. Rather than considering this
problem directly, the work here focuses on the model problem of a
charge coupled to a massless scalar field in a fixed (though
arbitrary) background spacetime. Some of the methods needed are
first introduced in the context of Newtonian gravity in Sect.
\ref{Sect:Newtonian}. Various generalizations necessary to work in
the relativistic case are then discussed in Sect.
\ref{Sect:ScalarForces}. A general prescription for the self-force
and self-torque acting on an extended charge is derived there, along
with a non-perturbative notion for the effective field momentum.
These results are finally applied in Sect. \ref{Sect:Apps} to obtain
the equations of motion satisfied by charges much smaller than any
significant timescale or curvature radius in the problem.

\section{Newtonian self-interaction}\label{Sect:Newtonian}

It is instructive to review the nonrelativistic self-force problem
before considering its generalizations. In a sense, this is trivial.
Net self-forces and self-torques acting on bodies in both Newtonian
gravity and ordinary electrostatics (with constant permittivity)
always vanish. Their existence is forbidden by Newton's third law.
Despite this, several important features in the analysis of this
simple problem persist even in the discussion of highly relativistic
systems. The methods used in this section are more complicated than
immediately necessary, although their unusual features are essential
for subsequent generalizations.

Proving that Newtonian self-forces vanish first requires a precise
definition for the field that generates them. This is, of course,
meant to be the portion of the field produced by the object itself.
Consider a compact body with finite radius that interacts with the
external universe purely via Newtonian gravity. The total
gravitational potential $\phi$ is then determined by
\begin{equation}
  \nabla^2 \phi(\bfx,s) = 4 \pi \rho(\bfx,s) ,
  \label{FieldNewt}
\end{equation}
where $\rho(\bfx,s)$ represents the mass density at time $s$. This
equation can be systematically solved for all reasonable mass
distributions by introducing a symmetric Green function $G(\bfx,
\bfx') = G(\bfx', \bfx)$ satisfying
\begin{equation}
  \nabla^2 G(\bfx,\bfx') = 4 \pi \delta( \bfx,
  \bfx'). \label{GreenNewt}
\end{equation}
This equation has a unique solution if $G$ is assumed to vanish when
its arguments are infinitely separated. Once it is known, the
gravitational potential produced by any mass distribution is
straightforward to compute. Denote the region occupied by the body
in question at time $s$ by $\Sigma(s)$. It is then natural to let
the self-field be given by
\begin{equation}
  \phiS(\bfx,s) = \int_{\Sigma(s)}  \rho(\bfx',s)
  G (\bfx, \bfx') \rmd V' .
  \label{SelfFieldDefNewt}
\end{equation}
While this is a very common definition, it is not the only one. Some
authors use the term in a purely perturbative sense indicating a
difference between the total field with and without the body of
interest \cite{DetPoisson}. While useful in some specialized
contexts, interactions with external matter make it very difficult
to derive any general properties of these difference fields. The net
forces they generate do not necessarily vanish, for example. Those
associated with \eqref{SelfFieldDefNewt} do, so they are all that
will be considered here.

It is straightforward to explore the consequences of self-fields
with this form. As is typical with self-force or radiation reaction
problems, the focus will be on determining the bulk or
``macroscopic'' aspects of a body's motion. Intricate details of an
object's shape and internal composition are ignored as much as
possible. The hope is that there exist a small number of state
parameters that generically describe some interesting behavior in a
large class of compact bodies. The typical example of such a
parameter is the center-of-mass position $\bm{\gamma}(s)$. At least
in certain limits, this couples very weakly to an object's shape.
Most state variables that are typically considered can be derived
from a body's net linear and angular momenta. There is an important
reason for this. If the laws of motion which are to be derived are
as general as hoped, the physics used to obtain them should be
similarly generic. One might expect to make use of geometric
structures in the background space and their effects on the laws of
motion of arbitrary systems. The obvious examples derive from
results like momentum conservation that are associated with
underlying geometric symmetries.

The three-dimensional Euclidean space of Newtonian physics (as
traditionally formulated) admits a six-parameter family of Killing
fields. Given some closed system, each of these is associated with a
conserved quantity built from the total linear and angular momenta.
If the velocity field of the matter is denoted by $u^a$, the
conserved quantity associated with a Killing field $K^a$ has the
form
\begin{equation}
  \ItP_K^{\mathrm{tot}}  = \int_{\mathcal{M}} \rho u_a K^a \rmd V ,
  \label{PKtot}
\end{equation}
where $\mathcal{M}$ denotes the entire space. Translational Killing
fields generate components of the system's linear momentum, while
rotational Killing fields generate components of its angular
momentum. Consider only the behavior of a particular body with
finite radius. Its momenta are parameterized by an analog $\ItP_K$
of \eqref{PKtot} obtained by integrating $\rho u_a K^a$ over
$\Sigma(s) \subset \mathcal{M}$. Such quantities are not usually
conserved.

The time-dependence of each $\ItP_K$ is easily derived from the
standard equations of continuum mechanics. A mass distribution with
stress tensor $\Sigma_{ab} = \Sigma_{(ab)}$ generically satisfies
\begin{equation}
  \frac{\partial}{\partial s} ( \rho u_a) + \nabla_b ( \rho u_a u^b
  + \Sigma_{a}{}^{b} ) = - \rho \nabla_a \phi . \label{ContMech}
\end{equation}
This could partially describe the dynamics of some elastic solid. It
is exactly Euler's equation for a perfect fluid if the stress tensor
is proportional to the metric. Regardless, the component of a body's
momentum generated by a Killing field $K^a$ must evolve according to
\begin{equation}
  \dot{\ItP}_{K} = \frac{\rmd \ItP_{K}}{\rmd s}  = - \int_\Sigma \rho \Lie_K \phi . \label{PDotNewt}
\end{equation}
The stress tensor does not appear explicitly in this equation, so
the net forces and torques are approximately independent of the type
of material under consideration. This independence is not exact
because $\Sigma_{ab}$ is still present implicitly in the equations
governing changes in the mass density.

The interpretation of the scalars $\ItP_K$ as components of momenta
can be made more clear by directly introducing such objects as
tensors at the mass center $\bm{\gamma}(s)$. Using standard
definitions for the linear momentum $p_a$ and angular momentum
$S_a$,
\begin{equation}
  \ItP_K = p_a K^a + \frac{1}{2} \epsilon_{a b c} S^a \nabla^{[b}
  K^{c]} = p_a K^a + \frac{1}{2} S_{a b} \nabla^{[a}
  K^{b]} . \label{PDefNewt}
\end{equation}
The second equality introduces the dual $S_{ab} = \epsilon_{abc}
S^c$ to the usual angular momentum vector. This is the more
fundamental quantity in the relativistic case, although the two
objects are interchangeable in three dimensions. Appropriate choices
for the Killing field in \eqref{PDefNewt} can be used to extract any
component of the linear or angular momenta. As an example, the
translational vector field fixed by setting $\nabla_a K_b=0$ and
$K_a = \nabla_a x$ at some $\bm{\gamma}(s)$ would recover the
$x$-component of $p^a(s)$. Complete knowledge of the family $\ItP_K$
for all possible Killing fields is equivalent to that of $p^a$ and
$S^{ab}$. It is more than sufficient to extract the center-of-mass
motion.

Now consider only the self-field's effect on the momenta. Combining
\eqref{SelfFieldDefNewt} with \eqref{PDotNewt} shows that
\begin{equation}
  \dot{\ItP}^\rmSelf_K =  - \frac{1}{2} \int_{\Sigma} \rmd V \! \int_{\Sigma} \rmd
  V' \rho \rho' \Lie_K G(\bfx, \bfx') ,
  \label{PDotSelfNewt}
\end{equation}
where $\rho = \rho(\bfx,s)$ and $\rho' = \rho(\bfx',s)$. Lie
derivatives of two-point functions are defined to act independently
on both of the subject's arguments, so
\begin{equation}
  \Lie_K G(\bfx, \bfx') = K^a(\bfx) \nabla_a G(\bfx, \bfx') + K^{a'}(\bfx') \nabla_{a'} G(\bfx, \bfx') .
\end{equation}
The derivation of \eqref{PDotSelfNewt} effectively replaced $K^a
\nabla_a G$ with $\Lie_K G/2$ by commuting integrals. While the
mathematical justification for this is clear, it is interesting to
mention its physical significance. The operation effectively
averages ``action-reaction pairs'' in the sense of Newton's third
law. It says that bulk self-field effects arise only if there are
imbalances between the forces exerted by (say) mass in $\rmd V$ on
mass in $\rmd V'$ versus the reverse. This is exactly what would be
expected from intuitive considerations. Proving that Newtonian
self-forces and self-torques vanish now requires only one more
ingredient.

The Green function adopted here has been fixed by choosing it to
vanish at infinity. The simplicity of this boundary condition
together with the form of \eqref{GreenNewt} implies that $G$ can
only depend the distance between its arguments. In anticipation of
later generalizations, it may be thought of purely as a function of
Synge's world function $\sigma(\bfx, \bfx')$. This biscalar returns
one half of the geodesic distance between its arguments
\cite{PoissonRev,Synge}. Translating or rotating any two points by
equivalent amounts does not change the distance between them, so
\begin{equation}
  \Lie_K \sigma(\bfx, \bfx') = 0
  \label{LieSigNewt}
\end{equation}
for any Killing field $K^a$. Substituting this result into
\eqref{PDotSelfNewt} immediately shows that $\dot{\ItP}^\rmSelf_K =
0$. This is the desired result: compact objects do not experience
any self-force or self-torque in Newtonian gravity. It is a
statement completely independent of a body's shape or detailed
structure. Essentially all that was used was the translational and
rotational invariance of the Green function and the generic equation
of motion \eqref{ContMech}. Invariance under translational Killing
fields is equivalent to the weak form of Newton's third law.
Supplementing it with the rotational invariance recovers the strong
form. An almost identical calculation leads to similar conclusions
in ordinary electrostatics (with similarly simple boundary
conditions) and many other theories. While this result could have
been derived more directly, many aspects of the method presented
here can now be generalized to analyze fully relativistic systems.
Despite an apparent reliance on the symmetries of Euclidean space,
geometric objects can be defined that allow similar manipulations
even in spacetimes admitting no Killing vectors at all.

Before discussing this, it should first be noted that there is a
complementary method of understanding the Newtonian self-force
problem. The approach just described effectively sums up the forces
acting inside an extended body. Identical conclusions can also be
obtained purely from the distant behavior of the gravitational
field. Combining \eqref{FieldNewt} and \eqref{PDotNewt},
\begin{equation}
  \dot{\ItP}_K = - \frac{1}{4\pi} \oint_{\partial \Sigma} [
  \nabla^a \phi \Lie_K \phi - \frac{1}{2} \nabla^b \phi \nabla_b (
  K^a \phi) ] \rmd S_a \label{SurfaceNewt}
\end{equation}
The effect of the self-field may be found by substituting $\phi
\rightarrow \phi^\rmSelf$ in this equation. The surface integral can
then be evaluated over closed surfaces outside of $\partial \Sigma$,
if desired. It is convenient to consider spheres extending to
infinity. The potentials are harmonic functions in this region, so
they must fall off at least as fast as $1/r$ as $r \rightarrow
\infty$. Any term that does decrease this slowly cannot have any
angular dependence. These two facts together with the properties of
the Killing fields show that all surface integrals like
\eqref{SurfaceNewt} must vanish. It follows that
$\dot{\ItP}_K^\rmSelf = 0$, as expected. Similar (though much more
complicated) derivations can be applied in the relativistic
self-force problem, although most of the discussion below takes the
more local viewpoint embodied by the derivation of
\eqref{PDotSelfNewt}.

\section{Relativistic scalar fields}\label{Sect:ScalarForces}

The discussion just presented suggests that self-forces and
self-torques could arise from local asymmetries in a field's
underlying Green function. Indeed, very small changes in the
statement of the Newtonian self-interaction problem allows for the
existence of nontrivial self-forces. Replacing the metric in the
field equation \eqref{FieldNewt} with one that isn't maximally
symmetric easily accomplishes this, for example. Using an elliptic
differential operator constructed from non-geometric objects can
have a similar effect. While \eqref{PDotSelfNewt} should not
necessarily be blindly applied in such cases, it is clear that
significant self-forces may arise.

These sorts of modifications are physically relevant in several
contexts. Static systems involving a curved spacetime are often
simplified with the use of a dimensional reduction procedure.
Laplace operators constructed from non-Euclidean metrics then arise
naturally in the field equations. Another interesting case is that
of ordinary electrostatics in the presence of dielectric materials
\cite{BurkoDielectric}. Even though the underlying space is very
simple, the field equation needn't be invariant under translations
or rotations. Both of these systems allow self-fields to strongly
affect the evolution of a body's net linear and angular momenta.
Relativistic extended bodies moving in curved spacetimes experience
very similar effects.

Other mechanisms are also at work, however. The transition to a
relativistic system involves fundamental changes in the character of
the fields. Mathematically, they usually shift from solutions of an
elliptic to a hyperbolic differential equation. This has several
physical consequences. Newtonian potentials are uniquely determined
by the instantaneous distribution of mass in the universe, for
example. Relativistic potentials are not. The past history of a
system is effectively remembered by the field in a complicated way.
It acquires its own degrees of freedom, and may transport energy and
momentum at finite speed. These differences lead to the importance
of radiation reaction and tail effects in self-force problems. While
not completely independent of each other, the three mechanisms just
described can all lead to significant self-forces. A useful model
system in which to illustrate these statements consists of a finite
material body interacting with a scalar field $\Phi(x)$. The
background spacetime will be assumed fixed and well-behaved in a
neighborhood of the body's worldtube $W$. Effects related to
gravitational self-interaction will be ignored here.


The type of scalar field chosen is not particularly important as
long as its field equation is linear. Still, some steps carried out
below can be copied over from previous work if it is assumed that
$\Phi$ is a massless minimally-coupled field satisfying
\begin{equation}
  \Box \Phi(x) = - 4 \pi \rho(x) .
    \label{KleinGordon}
\end{equation}
$\rho$ represents the scalar charge density in this equation. It is
straightforward to allow for a finite field mass or curvature
coupling, although this is an unnecessary complication. In the
Newtonian case, the field equation was needed mainly to define a
Green function. The same is true here. Let
\begin{equation}
    \Box G(x,x') = - 4 \pi \delta(x,x').
    \label{GreenDef}
\end{equation}
Solving this equation requires that certain boundary conditions be
imposed. The physical self-field will be defined by those associated
with the retarded Green function $\Gret$:
\begin{equation}
  \Phi^\rmSelf(x) = \int_W \rho(x') \Gret(x,x') \rmd V' .
\end{equation}
By construction, only points on the worldtube lying in the causal
past of $x$ contribute to this integral.

The scalar self-force problem now asks how such a field affects the
bulk motion of the charge that sources it. As in the Newtonian case,
it is reasonable to proceed by computing shifts in the body's
momenta. Appropriate analogs of the scalars \eqref{PKtot} take the
form
\begin{equation}
  \ItP_\xi(s) = \int_{\Sigma(s)} T^{a}{}_{b} \xi^b \rmd S_a ,
  \label{PDefInt}
\end{equation}
where $\xi^a$ is an as-yet unspecified vector field, $T^{ab}$ the
body's stress-energy tensor, and $\Sigma(s)$ some spacelike
hypersurface. The family of all such hypersurfaces is assumed to
foliate the worldtube $W$. It is not generally possible to choose
the generating vector fields in \eqref{PDefInt} to be Killing.
Despite this, there should be some sense in which they come as close
as possible to this ideal. A set of approximate Killing fields
suggested in \cite{HarteSyms} will be adopted here. These exactly
satisfy
\begin{equation}
\LieX g_{a b}|_\Gamma = \nabla_a \LieX g_{bc}|_\Gamma = 0,
\label{LieGKF}
\end{equation}
where $\Gamma$ is a preferred timelike worldline involved in their
construction. They are completely fixed throughout $W$ by the values
of $\xi^a$ and $\nabla_a \xi_b = \nabla_{[a} \xi_{b]}$ at any point
on this worldline. Each choice of initial data in this form defines
a unique generalized Killing field\footnote{The approximate
symmetries defined in \cite{HarteSyms} took the form of vector
fields with the property (among many others) that $\nabla_a \LieX
g_{b c} |_{\Gamma} = 0$. These were called generalized affine
collineations, or GACs. Some satisfy $\LieX g_{a b}|_\Gamma =0$, and
it is this subset of vector fields that properly generalize the
Killing fields. They were referred to as Killing-type GACs before,
although we now shorten this to generalized Killing fields (GKFs).
They are all that will be used here.} (GKF) $\xi^a$. Any genuine
Killing fields that may exist are in this class. The set of all GKFs
form a generalization $GP$ of the Poincar\'{e} group. Like the
standard Poincar\'{e} group, it has ten dimensions in four
dimensional spacetimes.

This summary of suggests an analog to \eqref{PDefNewt}. If $p^a(s)$
and $S_{a b} = S_{[ab]}(s)$ are the body's linear and angular
momenta represented as tensors at $\gamma(s) = \Sigma(s) \cap
\Gamma$, a relation of the following form should exist:
\begin{equation}
  \ItP_\xi = p^a \xi_a + \frac{1}{2} S^{a b}
  \nabla_{[a} \xi_{b]} .
  \label{PDef}
\end{equation}
This may be taken as a definition. The resulting momenta are exactly
those suggested by Dixon as being particularly useful for
understanding the mechanics of extended bodies in curved spacetimes
\cite{HarteSyms,Dix70a, Dix74, Dix79}. Like the GKFs, they depend on
both $\Gamma$ and $\Sigma$. These objects will be assumed to be
fixed using center-of-mass conditions \cite{EhlRud,CM}. Varying over
all possible GKFs, $\ItP_\xi$ becomes a map from $GP \times
\mathbb{R} \rightarrow \mathbb{R}$. Knowledge of its behavior is
completely equivalent to knowledge of $p_a$ and $S_{ab}$. These
quantities are sufficient to determine a body's mass, spin,
center-of-mass worldline, and so on. This includes almost all of the
local parameters typically computed in self-force problems. There is
also a sense in which it extracts all of the information that can be
recovered purely from stress-energy conservation \cite{Dix74}.

Rates of change of the scalar momenta $\ItP_\xi$ are easily related
to more standard definitions for forces and torques. As discussed in
\cite{HarteSyms}, the GKFs satisfy Killing transport equations on
$\Gamma$. It then follows from \eqref{PDef} that
\begin{equation}
  \rmd \ItP_\xi / \rmd s = (\dot{p}^a - \frac{1}{2}
  S^{bc} \dot{\gamma}^d R_{bcd}{}^a) \xi_a + \frac{1}{2} (
  \dot{S}^{ab} - 2 p^{[a} \dot{\gamma}^{b]} ) \nabla_{[a}
  \xi_{b]} . \label{pDotandPDot}
\end{equation}
Knowing the left-hand side allows the instantaneous force $F^a$ and
torque $N_{ab} = N_{[ab]}$ to be extracted. These objects are
typically defined such that \cite{Dix70a}
\begin{eqnarray}
  \dot{p}^a = F^a + \frac{1}{2} S^{bc} \dot{\gamma}^d R_{bcd}{}^a
  \label{Force}
  \\
  \dot{S}_{ab} = N_{ab} +  2 p_{[a} \dot{\gamma}_{b]} .
  \label{Torque}
\end{eqnarray}
Note that the Papapetrou equations hold if $F^a = N_{ab} = \rmd
\ItP_\xi/ \rmd s =0$.

Many consequences of $\Phi^\rmSelf$ cannot be determined from its
direct effect on the momenta. Self-fields strongly influence the
equilibrium shapes (and therefore the higher multipole moments) of
highly charged objects, help resist tidal deformations, perturb
distant matter, and so on. These phenomena usually couple very
weakly to $\ItP_\xi$. They are effectively ignored by the current
formalism. This does not necessarily mean that they are negligible
compared to the effects considered here. Especially in the context
of gravitational self-forces, the self-forces defined in the present
manner may be comparable to shifts in the external field due to
perturbations of distant masses. This is illustrated explicitly in
\cite{PfenningPoisson}, and is a standard problem. It can usually be
made less severe in the scalar and electromagnetic self-force
problems, so we will ignore it. The methods introduced here can at
least be used to simplify a significant portion of the overall
problem of motion.

The time-dependence of the momenta $\ItP_\xi$ follows from
stress-energy conservation. If the only long-range field other than
gravity is $\Phi$, this requires that
\begin{equation}
  \nabla_a (T^{ab} + t^{ab} ) =0,
  \label{StressCons1}
\end{equation}
where the stress-energy tensor of the scalar field is
\begin{equation}
  t^{ab} = \frac{1}{4 \pi} ( \nabla^a \Phi \nabla^b \Phi - \frac{1}{2} g^{ab}
  \nabla_c \Phi \nabla^c \Phi ) .
  \label{StressScalar}
\end{equation}
Combining these expressions with \eqref{KleinGordon} gives
\begin{equation}
  \nabla_a T^{a}{}_{b} = \rho \nabla_b \Phi.
\end{equation}
The Newtonian limit of this equation is essentially identical to
\eqref{ContMech}. Its right-hand side represents the force density
exerted by the scalar field on the matter distribution.

The evolution of the momenta is more convenient to analyze in terms
of finite differences
\begin{equation}
  \delta \ItP_\xi (s_2,s_1) = \ItP_\xi(s_2) - \ItP_\xi(s_1)
\end{equation}
rather than instantaneous rates of change. These might represent
changes in a body's mass or spin over the time interval $(s_1 ,
s_2)$. Suppose that $s_2 > s_1$, and that $s$ increases
monotonically as $\gamma(s)$ extends into the future. If $\Omega =
\Omega(s_1,s_2)$ denotes the portion of $W$ lying between the
hypersurfaces $\Sigma(s_1)$ and $\Sigma(s_2)$, and
$T^{ab}|_{\partial W} =0$, Gauss' theorem shows that
\begin{equation}
  \delta \ItP_\xi = \int_{\Omega} \left( \frac{1}{2} T^{ab}
  \LieX g_{ab} + \rho \LieX \Phi  \right) \rmd V.
  \label{DP}
\end{equation}
The first term here represents gravitational force and torque. It
exists regardless of whether any scalar field is present. On each
$\Sigma(s)$, this portion of the integrand can be shown to be
equivalent to a Christoffel symbol contracted with $T^{ab}$ in a
normal coordinate system based at $\gamma(s)$ \cite{Dix79}. If the
background geometry varies slowly throughout the body (both
spatially and temporally), this gravitational term can be expanded
in terms of the multipole moments of the stress-energy tensor. The
lowest order contribution comes from the quadrupole, and is
relatively simple to take into account. Detailed examples of this
exist in the literature \cite{HarteQuadrupole, ItalianQuadrupole}.

This leaves only the scalar field's contribution to the momentum
shift. It may be split into two parts. First consider the portion
due to $\Phi^\rmSelf$. Let
\begin{equation}
  \delta \ItP^\rmSelf_\xi = \int_{\Omega} \rho \LieX
  \Phi_{\rmSelf} \rmd V  .
  \label{DPSelf}
\end{equation}
The remaining (external) component of the scalar field usually
varies slowly inside the body. Its contribution to the motion may
therefore be evaluated using another multipole expansion. Similar
methods cannot be directly used to understand the self-force. As it
stands, $\Phi^\rmSelf$ almost always varies rapidly over scales
comparable to the body's proper radius. Successive terms in a
multipole expansion of \eqref{DPSelf} would therefore fail to
decrease in magnitude. Such a series would not be useful.

The form of the self-force simplifies if another split is made. Let
$T^{-}_\Omega$ denote the portion of $W$ lying in the exclusive past
of $\Omega$; i.e. $T^{-}_\Omega = (J^-[\Omega] \setminus \Omega)
\cap W$ in the notation of \cite{Wald}. Defining the retarded field
sourced by charge in an arbitrary region $\Lambda$ by
\begin{equation}
    \PhiRet [\Lambda] = \int_{\Lambda} \rho'
    \Gret \rmd V' , \label{PhiRetArb}
\end{equation}
it is trivially true that
\begin{equation}
  \Phi^\rmSelf(x) = \PhiRet[W] = \PhiRet[\Omega] + \PhiRet[
  T^{-}_\Omega ]
  \label{FieldSplit}
\end{equation}
for any $x \in \Omega$. The second term here will be left as-is for
now. Its contribution to the self-force is reasonably well-behaved
even for a $\delta$-function source. The field due to charge in
$\Omega$ is more interesting. This is where most of the Coulomb and
other quickly-varying components of the self-field arise.

Forces and torques exerted by $\PhiRet[\Omega]$ can be simplified by
introducing regular and singular Green functions $\GR$ and $\GS$.
For now, these objects will only be required to satisfy
\begin{equation}
  \Gret = \GR + \GS
  \label{GreenSum}
\end{equation}
and the reciprocity relation $\GS(x,x') = \GS(x',x)$. It will later
be useful to also suppose that
\begin{equation}
  \Box \GR = 0 .
  \label{GradDef}
\end{equation}
This contradicts \eqref{GreenDef} -- which applies for both $\Gret$
and $\GS$ -- so it is something of a misnomer to call $\GR$ a Green
function. It is common to do so, however, and this practice will be
followed here. These properties have been chosen so that $\GS$ is as
close to a Newtonian Green function as possible. This presumably
minimizes its influence on the body's overall motion.

Many propagators with these properties exist, however. Perhaps the
simplest derives from using a $\GS$ with the form
\begin{equation}
  G_{\mathrm{S,D}} = \frac{1}{2} (\Gret + \Gadv), \label{GSingular}
\end{equation}
where $\Gadv$ is the advanced Green function. The regular or
radiative Green function derived from this choice using
\eqref{GreenSum} was central to Dirac's classical electron model
\cite{Dirac}. It will be referred to here as the Dirac Green
function. Another possibility is to use the construction given by
Detweiler and Whiting in connection with point particle self-force
regularization in curved spacetimes
\cite{DetWhiting,PoissonRev,WhitingSing}. Regardless of their
specific definitions, the names given to these objects derive from
their connection to point particle self-fields. The linearity of the
field equation and \eqref{GreenSum} suggest that such fields may be
split into singular and regular components respectively sourced by
$\GS$ and $\GR$. For a point particle, the portion derived from the
singular Green function diverges on its worldline. The remainder of
the self-field remains bounded even at the source's location. It is
typically associated with radiation. Self-forces are often thought
of (somewhat incompletely) as the local reaction to emitted
radiation, so one would expect most of the self-interaction to arise
from fields associated with $\GR$. Note that neither of the Green
functions introduced here lead to any singular behavior for
well-behaved extended charge distributions.

Regardless of which specific choices are made for $\GR$ and $\GS$, a
relativistic analog of \eqref{PDotSelfNewt} is easily derived. Pair
averaging is only meaningful for fields derived from Green functions
that are symmetric in their arguments. This is one of the defining
properties of $\GS$, so the averaging will only be applied on the
singular portion of the self-field. It is then straightforward to
show that \eqref{DPSelf} is equivalent to
\begin{eqnarray}
  \delta \ItP^\rmSelf_\xi = \int_\Omega \rmd V \rho \Big[ \LieX ( \PhiR [\Omega] + \PhiRet[ T^{-}_\Omega ] ) + \frac{1}{2} \int_{\Omega} \rmd V'
  \rho' \LieX \GS \Big] .
    \label{DPFinalScalar}
\end{eqnarray}
$\PhiR [\Lambda]$ is defined here by analogy to \eqref{PhiRetArb}.
The regular self-field affecting the momentum shift in this equation
only depends on charge in $\Omega$. There are therefore no
conceptual obstacles to adopting a $\GR$ with support in the
chronological future of the field point.

This freedom has an undesirable consequence. $\delta
\ItP^\rmSelf_\xi$ is trivially interpreted as the time average of
$\dot{\ItP}^\rmSelf_\xi$. The instantaneous force or torque is
expected to depend only on the properties of the physical system and
the choice of GKF. It is not affected by arbitrary parameters like
$s_1$ and $s_2$. Individual terms on the right-hand side of
\eqref{DPFinalScalar} might be expected to share this property. They
do not. The various fields there are all derived from sources with
sharp temporal boundaries. Their behavior always changes abruptly
near these regions. This has no physical significance. The total
motion is unaffected (as it must be), although it makes the
interpretation of the various self-force contributions more
difficult. Such effects can be separated out explicitly. Rewriting
\eqref{DPFinalScalar},
\begin{eqnarray}
  \delta \ItP^\rmSelf_\xi =& \int_{\Omega} \rmd V \rho \Big( \LieX
  \PhiR[W] + \frac{1}{2} \int_W \rmd V' \rho' \LieX \GS \nonumber
  \\
  & ~ + \LieX \PhiS[ W \setminus \Omega] - \frac{1}{2} \int_{W \setminus \Omega} \rmd V'
   \rho' \LieX \GS \Big).
  \label{DPFinalScalar2}
\end{eqnarray}
The first two terms in parentheses here only depend on properties of
the physical system. The remaining quantities are different. They
are directly linked to the choice of $\Omega$. No matter how simple
the charge distribution and external fields may be, this portion of
the integrand always changes character near $\Sigma(s_1)$ and
$\Sigma(s_2)$. Its contributions to the self-force would be
simplified if there was a sense in which they only contributed to
the integral near these hypersurfaces.

This is accomplished by adding an additional axiom to those
constraining the singular and regular Green functions. Although the
specific Dirac form \eqref{GSingular} for $\GS$ is very simple, it
generically has support in the entire causal past and future of any
field point. Computing terms like $\LieX \PhiS[W \setminus \Omega]$
in \eqref{DPFinalScalar2} would then require knowing the entire past
and future history of the system. There is no region in causal
contact with $W$ where this quantity would be expected to vanish.
Suppose that another singular Green function is chosen that always
vanishes whenever its arguments are timelike-separated. This is true
of \eqref{GSingular} only in flat spacetime. More generally, this
assumption together with the original axioms constraining $\GS$ and
$\GR$ uniquely specify the aforementioned Detweiler-Whiting Green
functions \cite{DetWhiting,PoissonRev,WhitingSing}.

Before specifying these objects explicitly, it is first useful to
review the Hadamard decomposition for the singular Green function
defined in \eqref{GSingular}. This has the form
\begin{equation}
  G_{\mathrm{S,D}} (x,x') = \frac{1}{2} [ \Delta^{1/2} \delta(\sigma) + V
  \Theta(-\sigma) ]_{(x,x')} .
  \label{Hadamard}
\end{equation}
There are two distinct contributions here. One -- familiar from the
study of massless fields in $3+1$ dimensional Minkowski spacetime --
is concentrated entirely on the light cones of the field point $x$.
The biscalar coefficient $\Delta$ is known as the van Vleck
determinant. It may be expressed in terms of the first two
derivatives of world function $\sigma$ via \cite{PoissonRev}
\begin{equation}
  \Delta(x,x') = \frac{\det[ - \nabla_a \nabla_{a'} \sigma(x,x') ] }{ \sqrt{-g} \sqrt{-g'} }
  . \label{VanVleck}
\end{equation}
In all reasonable cases of interest here, this is smooth, positive,
and reduces to unity as $x \rightarrow x'$. Its first derivatives
also vanish at coincidence. The second (tail) term in
\eqref{Hadamard} takes into account that disturbances in the field
do not necessarily propagate only on null rays. $V(x,x')$ depends on
the details of the spacetime, and is almost always nonzero. It does
remain smooth, however. While $V$ is usually difficult to find, its
coincidence limit is known to be \cite{PoissonRev}:
\begin{equation}
    \lim_{x' \rightarrow x} V(x,x') = \frac{1}{12} R(x) .
\end{equation}
Both $\Delta$ and $V$ are symmetric in their arguments.

Using all of these definitions, the Detweiler-Whiting singular Green
function may be shown to have the form
\cite{DetWhiting,PoissonRev,WhitingSing}
\begin{equation}
    \GSDW = G_{\mathrm{S,D}} - \frac{1}{2} V = \frac{1}{2} [ \Delta^{1/2} \delta( \sigma ) - V \Theta(
    \sigma ) ] .
    \label{GSingularDW}
\end{equation}
This generically has support everywhere but in the chronological
past or future of either of its arguments. The regular
Detweiler-Whiting Green function obtained from \eqref{GreenSum} has
support everywhere except inside the future null cone of each field
point. These propagators are therefore acausal. All derivations here
have started by expanding retarded Green functions, so this has no
unphysical consequences. Adopting the Detweiler-Whiting Green
functions greatly simplifies the interpretation of
\eqref{DPFinalScalar2}. Each term in that equation might have
initially appeared to involve knowledge of the body's behavior in
the infinite future. While the sum of all such contributions cancels
out, it is not immediately obvious how this occurs. Setting $\GS =
\GSDW$ largely removes this problem. Knowledge of the system then
appears to be required only out to times of order the body's
diameter beyond $s_2$. Such contributions still do not have direct
physical consequences, although they are now much simpler to control
and understand.

A more concrete advantage of these special Green functions is that
the meaning of the Detweiler-Whiting axiom can now be clarified.
This states that the self-field derived from $\GSDW$ exerts no force
on a point particle \cite{DetWhiting,PoissonRev,WhitingSing}.
Equations of motion obtained with this assumption are identical to
those appearing in all other treatments of point particle motion. It
is therefore interesting to see how well it applies for a finite
extended body. It cannot be exact, as it is known that the singular
self-field contributes an effective mass to extended charges. This
is a consequence of the fact that accelerating a particle requires
accelerating both its matter and field components. It is a somewhat
trivial effect in the sense that masses like those defined by
\begin{equation}
  m = \sqrt{ - p^a p_a} \label{MassDef}
\end{equation}
are rarely measured directly. Doing so would require detailed
knowledge of an object's stress-energy tensor. More realistically,
inertial masses are usually measured by observing an object's motion
under the application of known external forces. This method would
recover a mass that included contributions from both $p^a$ and the
self-field. The interesting question is therefore whether there are
any effects on a body's motion induced by its singular self-field
that cannot be attributed merely to a (possibly time-dependent) mass
shift. Calculations using perturbation theory in flat spacetime
electromagnetism have found such phenomena even in cases where the
Detweiler-Whiting axiom applied to a finite charge would imply the
point particle equations of motion \cite{HarteEM}. Interestingly,
the methods introduced so far allow significant insight to be gained
into this result without the use of any approximations.

First restrict attention to momentum shifts over times $\delta s$
longer than the body's light-crossing time $D$. More precisely,
assume that every point in $\Sigma(s_2)$ is timelike-separated from
every point in $\Sigma(s_1)$. Setting $\GS = G_{\mathrm{S,DW}}$, the
last two terms in \eqref{DPFinalScalar2} may then be associated
entirely with the boundary caps of $\Omega$. Their contribution to
the self-force and self-torque has the form $\mathcal{E}_\xi(s_1) -
\mathcal{E}_\xi(s_2)$, where
\begin{eqnarray}
  \mathcal{E}_\xi =& \frac{1}{2} \Big( \int_{\Sigma^{+}} \rho \LieX \Phi^{\mathrm{S,DW}} [
  \Sigma^{-} ] \rmd V - \int_{\Sigma^{-}} \rho \LieX \Phi^{\mathrm{S,DW}} [
  \Sigma^{+} ] \rmd V  \Big) . \label{EDef}
\end{eqnarray}
If $\mathcal{E}_\xi = \mathcal{E}_\xi(s)$, the two regions of
integration in this equation bisect the body's worldtube.
$\Sigma^{+}(s)$ denotes the portion of $W$ to the future of
$\Sigma(s)$, while $\Sigma^{-}(s)$ represents the volume to its
past. Both of these domains are unbounded, although the definition
of the singular Green function used here effectively restricts them
to small volumes extending over time intervals of order $D$ away
from $\Sigma(s)$.

Combining \eqref{DPFinalScalar2} and \eqref{EDef}, all compact
charge distributions are found to satisfy
\begin{eqnarray}
  \delta( \ItP^\rmSelf_\xi + \mathcal{E}_\xi) =
  \int_\Omega \rmd V \rho \Big( \LieX \Phi_{\mathrm{R,DW}}[W] + \frac{1}{2} \int_W \rmd V' \rho' \LieX G_{\mathrm{S,DW}} \Big) .
  \label{DPFinalScalarAv}
\end{eqnarray}
This result is exact at least as long as $\delta s$ is not too
small. Considering differences in the momenta between three times
$s_1$, $s_2$, and $s_3$ satisfying $s_3-s_1 \gg D$ and $s_2 -s_1 \gg
D$ shows that it is actually correct for any time interval. It is
therefore possible to consider an instantaneous form of
\eqref{DPFinalScalarAv}. Restoring the forces directly due to the
geometry and the external scalar field,
\begin{eqnarray}
   \frac{\rmd}{\rmd s} ( \ItP_\xi + \mathcal{E}_\xi )  =& \int_\Sigma
    \Big[ \frac{1}{2} T^{ab} \LieX g_{ab} + \rho \LieX
  ( \Phi^{\mathrm{ext}} + \Phi_{\mathrm{R}}^\rmSelf ) \nonumber
  \\
  & \qquad ~ + \frac{1}{2} \int_W \rmd V' \rho \rho' \LieX \GS \Big] t^a \rmd S_a.
  \label{PBarAv}
\end{eqnarray}
$t^a$ is the time evolution vector field for the foliation $\{
\Sigma \}$. It is also implicit here that $\Phi_{\mathrm{R}}^\rmSelf
= \Phi_{\mathrm{R}}[W]$ and $\Phi^{\mathrm{ext}} = \Phi -
\Phi^\rmSelf$. This result holds for any $\GS$ and $\GR$. In
general, though, individual terms will require knowledge of the
system into the infinite future. It is only when the
Detweiler-Whiting Green functions are used that this dependence is
restricted to small times of order $D$.

The third term on the right-hand side of \eqref{PBarAv} is expected
from the Detweiler-Whiting axiom generalized for an extended body.
There are two corrections to this. As already discussed, the term
involving $\LieX \GS$ arises from averaging action-reaction pairs in
the sense of Newton's third law. It vanishes identically in
Minkowski and de Sitter spacetimes. The calculations in Sec.
\ref{Sect:Apps} also suggest that it generally contributes very
little to the motion of a charge that's sufficiently small compared
to the curvature scales of the background geometry.

The other interesting term in \eqref{PBarAv} involves
$\mathcal{E}_\xi$. Its presence suggests that there is a sense in
which the momenta derived from \eqref{PDefInt} are incomplete.
Extended charges seem to respond as though they had effective
momenta
\begin{equation}
  \hat{\ItP}_\xi = \ItP_\xi + \mathcal{E}_\xi. \label{PRen}
\end{equation}
The singular self-field is effectively conservative up to the term
involving $\LieX \GS$ in \eqref{PBarAv}. Comparing these
renormalized momenta with \eqref{EDef} and an expression analogous
to \eqref{PDef} can be used to define $\hat{p}_a$ and
$\hat{S}_{ab}$. Differences with their unhatted counterparts can be
interpreted as being due to the momenta of the singular component of
a particle's self-field. It straightforward to verify this
identification in stationary systems. Assuming that all relevant
quantities are time-symmetric about some $\Sigma$,
\begin{equation}
  \mathcal{E}_\xi = - \frac{1}{2} \int_{\Sigma} \rho \PhiS[W] \xi^a \rmd S_a
  .
\end{equation}
This is strongly reminiscent of the expression for a system's
self-energy. For a body in geodesic motion in flat spacetime, it has
the explicit form
\begin{equation}
  \mathcal{E}_\xi = \frac{1}{2} \int \rmd^3 \mathbf{r}  \rmd^3  \mathbf{r}'
  \rho(\mathbf{r}) \rho (\mathbf{r}') \left( \frac{ \dot{\gamma}^a \left[ \xi_a(\gamma) - r^i \nabla_a \xi_i(\gamma) \right] }{ |\mathbf{r} - \mathbf{r}'| } \right) .
\end{equation}
Minkowski coordinates have been used here in the obvious way. The
worldline defined by $\mathbf{r} =0$ corresponds to one used to
construct the GKFs. The net effect of $\mathcal{E}_\xi$ on
$\hat{p}_a$ in this case is to add to the bare mass $m$ a term equal
to the (singular component of the) particle's self-energy. The
effective angular momentum may also be changed by $\mathcal{E}_\xi$.
This shift is purely orbital in character, and vanishes when the
origin is placed at the center of the self-energy distribution. In
general, this point will not coincide with the center-of-mass
computed purely from $T^{ab}$.

It is typical to define a center-of-mass frame by demanding that
$\Gamma$ be chosen such that
\begin{equation}
  (p^a S_{ab} ) _\Gamma  =0 .
\end{equation}
Each $\Sigma(s)$ is to be formed from the set of all geodesics
passing through $\gamma(s)$ orthogonally to $p^a(s)$
\cite{EhlRud,CM}. The existence of an effective momentum here
suggests an alternative mass center $\hat{\Gamma}$ that could be
defined via
\begin{equation}
  (\hat{p}^a \hat{S}_{ab})_{\hat{\Gamma}} = 0, \label{GammaBar}
\end{equation}
along with an appropriate condition for a foliation $\{ \hat{\Sigma}
\}$. It is unclear which of these definitions more appropriately
captures some sense of a charge's average position. The laws of
motion \eqref{PBarAv} are simpler for the effective rather than the
bare momenta, so $\hat{\Gamma}$ should have simpler evolution
equations. This is not sufficient to justify assuming that the
resulting worldline is preferable, although it is suggestive. Of
course, there are many nontrivial cases where $\Gamma =
\hat{\Gamma}$, or where any differences are extremely small.

It is useful to compare the results obtained so far to those derived
(by very different methods) in \cite{HarteEM}. There, the motions of
a large class of extended electromagnetic charge distributions were
studied in flat spacetime. This was done by assuming that the
self-fields were derived from either retarded or regular Green
functions. The momenta naturally associated with electromagnetically
interacting bodies are more complicated than \eqref{PDefInt}
\cite{Dix70a,Dix74,Dix67}. There is an additional term involving the
electromagnetic field and current distribution that does not have an
analog in the scalar case considered here. Despite this, expressions
very similar to \eqref{PBarAv} might be expected to remain valid.
Any term involving a Lie derivative of the electromagnetic Green
function would vanish in flat spacetime, so the self-forces and
self-torques might be expected to involve only the regular
self-field and some analog of $\mathcal{E}_\xi$. Considerable
differences were found in \cite{HarteEM} between the regular and
retarded electromagnetic self-forces acting on the bare momenta.
This was true even after obvious mass rescalings were taken into
account. It therefore appears that changes in the self-momentum can
have a nontrivial effect on $\ItP_\xi$. Verifying this in the scalar
case would require detailed calculations that will not be attempted
here.

It was mentioned in Sect. \ref{Sect:Newtonian} that the motion of a
Newtonian mass can be determined either by locally analyzing its
internal forces or by studying the asymptotic structure of its
self-field. This is also true in the relativistic case. We have
focused on the local viewpoint so far. Alternatively, changes in the
matter's momenta may be viewed as arising from changes in the field
momenta. Using the standard stress-energy tensor
\eqref{StressScalar} for a scalar field, momenta can be associated
with $\Phi$ or $\Phi^\rmSelf$ just as they are with $T^{ab}$. Let
\begin{equation}
  \mathcal{U}_\xi = \int_\Sigma t^{ab} \xi_b \rmd S_a .
\end{equation}
The scalar field does not usually have compact support, so this
quantity depends on the details of $\Sigma$ outside of $W$.
Regardless, stress-energy conservation implies that changes in the
total momentum $\ItP_\xi + \mathcal{U}_\xi$ satisfy
\begin{equation}
  \frac{\rmd}{\rmd s} (\ItP_\xi + \mathcal{U}_\xi) = \frac{1}{2} \int_\Sigma (T^{ab} + t^{ab}) \LieX g_{ab} t^a \rmd S_a - \oint_{\partial \Sigma} t^{ab} \xi_a t^c \rmd S_{bc}.
\end{equation}
This is closely related to the Newtonian result \eqref{SurfaceNewt}.
The first integral measures the degree to which momentum fails to be
conserved in a curved spacetime. Such terms will usually become
negligible if the body is sufficiently small and $\Sigma$ does not
extend far outside of it. Unfortunately, the surface integral is
simplest to evaluate very far away from $W$. Determining the optimal
balance between these two competing influences is not trivial. It is
also not simple to compute $\rmd \mathcal{U}_\xi/\rmd s$. These
difficulties are mentioned merely for completeness. They do not
arise in the local description used to derive \eqref{PBarAv}.

\section{Small charges}\label{Sect:Apps}

Self-forces affecting the motion of arbitrary bodies can be
extremely complicated. Internal oscillations might produce
``radiation rockets,'' for example. Such effects can exist even in
the absence of any external influences. There is little that can
usually be said about these phenomena without considering specific
models. It is therefore more typical to focus on self-interactions
affecting small systems close to some stable equilibrium. Making
this idea precise can be difficult. In general, standard radiation
reaction effects are recovered by restricting a body's spin, the
position of its ``center-of-charge'' with respect to its mass
center, the magnitude of its self-energy, speeds of internal
motions, and many other parameters \cite{HarteEM}. This procedure is
very complicated, so it is common to ignore at the outset all
effects related to a system's internal structure.

\subsection{Distributional sources}\label{Sect:Point Particles}

Na\"{\i}vely, one might try to do this by analyzing the behavior of
point charges. In the scalar case considered here, a charge density
could be chosen with the form
\begin{equation}
  \rho(x) = \int q(t) \delta(x,z(t)) \rmd t .
  \label{PointParticle}
\end{equation}
This represents a particle with charge $q(s)$ concentrated entirely
on a worldline parameterized by $z(s)$. For simplicity, it is
usually assumed that the dipole moment vanishes (meaning that $z =
\gamma$) and the charge remains constant. It is well-known that the
self-field of such a source diverges like $1/r$ in normal
coordinates centered on its worldline. Results like \eqref{DPSelf}
then appear to be meaningless. A possible reaction to this is that
point particles of the given type are unphysical. It therefore isn't
necessary for the standard laws of physics to be compatible with
them. Despite this, many authors have introduced special
regularization methods intended to force such objects into the
theory \cite{Dirac,QuinnWald,Quinn,DetWhiting,RosenthalMassField}.

In keeping with this tradition, it is interesting to discuss how
point particles can be fit into the current formalism. The required
assumption is surprisingly simple. First suppose that the self-force
and self-torque are to be derived from \eqref{DPFinalScalarAv}. As
it stands, this equation is not useful for a point particle.
$\mathcal{E}_\xi$ roughly involves the self-energy, so it diverges.
This problem may be removed by only working with the effective
momenta $\hat{\ItP}_\xi$ defined in \eqref{PRen}. If it is assumed
that these quantities are always finite and that the particle's
worldline satisfies \eqref{GammaBar}, standard results -- first
derived by Quinn \cite{Quinn} -- follow when
\begin{equation}
  \int_\Sigma \rmd S_a t^a  \int_W \rmd V' \rho  \rho' \LieX \GS  =0.
  \label{PointResult}
\end{equation}
The main intention of this section is to demonstrate that this
relation holds in all spacetimes smooth near the particle. The
Detweiler-Whiting Green functions will be adopted here.

Before proving \eqref{PointResult}, note that the crucial step --
assuming that $\hat{\ItP}_\xi$ is finite -- is very similar to a
standard mass renormalization procedure. In that case, the force on
a small charge is shown perturbatively to involve a term of the form
$(\mathrm{self\!-\!energy}) \times \ddot{\gamma}^a$
\cite{OriExtend,Thirring}. This effectively acts to shift the
particle's mass. Although the self-energy diverges as $D \rightarrow
0$ (with $q$ fixed), it is assumed that its combination with $m$ is
finite. Here, the self-momentum was identified non-perturbatively,
and can be absorbed into ``observable parameters'' at the outset. It
was never necessary to obtain explicit solutions of the field
equation.

Trying to derive self-forces and self-torques from
\eqref{DPFinalScalar} instead of \eqref{PBarAv} leads to a slightly
different point of view. The problematic self-energy
$\mathcal{E}_\xi$ was originally found to arise from the behavior of
$\PhiR[\Omega]$ and $\PhiRet[T^{-}_\Omega]$ very near $\Sigma(s_1)$
and $\Sigma(s_2)$. The relevant volume shrinks to zero in the point
particle case. It is therefore very easy in performing the various
required integrations to miss it entirely. There is sufficient
ambiguity that it isn't really clear that it should be there at all.
A somewhat carefree application of \eqref{DPFinalScalar} would find
that no renormalization was necessary at all in order to obtain
finite forces and torques on point particles. This is one kind of
selective ignorance. Assuming that $\hat{\ItP}_\xi$ is finite is
another. The latter point of view will be adopted here. A fully
consistent analysis would consider extended charge distributions
whose mass, charge, and radius all shrink to zero at appropriate
rates. This will be discussed in Sect. \ref{Sect:Scaling Limit}
below.

We now derive \eqref{PointResult} for a point charge. Assume that
$\dot{q} = 0$ and that the charge density is concentrated on a
center-of-mass worldline $\hat{\Gamma}$ satisfying \eqref{GammaBar}.
This is the worldline that will be used to construct the GKFs. For
notational convenience, hats will be omitted for the remainder of
this section. It is implicit that all momenta and mass centers are
associated with $\hat{\ItP}_\xi$. The problem then reduces to
evaluating
\begin{equation}
\lim_{s_1 \rightarrow s_2} \frac{1}{\delta s} \int_{s_1}^{s_2} \rmd
t \int_{-\infty}^{\infty} \rmd t' \LieX \GSDW(\gamma(t), \gamma(t'))
. \label{DPLieG}
\end{equation}
Given \eqref{GSingularDW}, the integrand here involves terms
proportional to $\Theta(\sigma)$, $\delta(\sigma)$, and
$\delta'(\sigma)$. The first of these is manifestly finite, and
scales like $(s_2 -s_1)^2$ in the limit $s_1 \rightarrow s_2$. It is
therefore irrelevant in \eqref{DPLieG}. The potentially interesting
quantities are
\begin{equation}
  \lim_{t' \rightarrow t} \left[ \frac{ \Delta^{1/2}
  \LieX \ln \Delta/2 - V \LieX \sigma }{ |\dot{\gamma}^{a'} \sigma_{a'} | }
  \right]
  \label{PointLimit1}
\end{equation}
and
\begin{equation}
  \lim_{t' \rightarrow t} \left[ \frac{1}{ |\dot{\gamma}^{a'}
  \sigma_{a'}
  | } \frac{ \partial }{ \partial t'} \left( \frac{
  \Delta^{1/2} \LieX \sigma}{ \dot{\gamma}^{b'} \sigma_{b'}
  } \right) \right] .
  \label{PointLimit2}
\end{equation}
The standard notation $\sigma_{a'} = \nabla_{a'} \sigma$ has been
used in these expressions. It generalizes in the obvious way for any
combination of primed and unprimed indices.

The two limits here are easily computed using the properties of GKFs
derived in \cite{HarteSyms}. First consider Lie derivatives of the
world function evaluated on two nearby points on $\Gamma$. These
obviously vanish when the points coincide. What is needed is an
estimate for precisely how fast they tend to zero as $t \rightarrow
t'$. It will be sufficient to note that on the center-of-mass
worldline, \eqref{LieGKF} can be used to show that
\begin{eqnarray}
  \xi^{a'} \simeq - \sigma^{a'}{}_{a} [\xi^a + X^b
  \nabla_b \xi^a + \frac{2}{3} X^b X^c R^{a}{}_{bcd} \xi^d] + \Or(X^3)
  .
\end{eqnarray}
$X_a = - \sigma_a(\gamma,\gamma')$ acts like a separation vector
between its two arguments. Using the antisymmetry of $\nabla_a
\xi_b$ on $\Gamma$ together with the well-known identity
\cite{PoissonRev,Synge}
\begin{equation}
  \sigma_a = \sigma^{a'} \sigma_{aa'}
\end{equation}
shows that $\LieX \sigma(\gamma,\gamma')$ decreases at least as fast
as $(t -t')^4$ as these times approach each other. It is clear that
$\dot{\gamma}^{a'} \sigma_{a'}$ scales like $(t-t')^1$ in the same
limit. These two relations are sufficient to show that
\eqref{PointLimit2} always vanishes.

Understanding the remaining limit \eqref{PointLimit1} requires
knowing how fast $\LieX \ln \Delta$ decreases as $t \rightarrow t'$.
It is shown in \cite{HarteSyms} that
\begin{equation}
  \LieX \ln \Delta = - H^{a'}{}_{a} ( \xi^b \sigma^{a}{}_{ba'} +
  \xi^{b'} \sigma^{a}{}_{a'b'} ),
\end{equation}
where
\begin{equation}
  H^{a'}{}_{a} = [ - \sigma^{a}{}_{a'} ]^{-1} .
\end{equation}
The ``-1'' on the right-hand side of this equation denotes a matrix
inverse. It is assumed here that $H^{a'}{}_{a}$ exists in all
regions of interest. It reduces to the identity when its arguments
coincide. A straightforward application of Synge's rule
\cite{PoissonRev,Synge} and other standard results of bitensor
analysis shows that both $\LieX \ln \Delta$ and its first covariant
derivatives vanish in a similar limit. Such Lie derivatives
therefore scale like $(t-t')^2$ as $t \rightarrow t'$. Deriving this
result actually does not require any properties of the GKFs. It
holds for all smooth vector fields $\xi^a$. This scaling relation
together with the previously-discussed one for $\LieX \sigma$ imply
that \eqref{PointLimit1} always vanishes. It follows that
\eqref{PointResult} holds, as originally claimed. This is
effectively equivalent to stating that the singular self-field
always satisfies Newton's third law in sufficiently small regions
near $\Gamma$. Although it did not require any external assumptions,
this result can be thought of as an effective renormalization of the
point particle self-field. The degree to which the generalized
Killing fields live up to their name has removed any singularities
that might have been expected to arise from the field sourced by
$\GS$. Note, however, that this procedure cannot be applied to
charges with arbitrary distributional structures. It would also fail
if the point charge was not concentrated on the same worldline used
to define the GKFs.

Equations of motion for a point charge can now be derived from the
behavior of its momenta. Assume that the body's stress-energy tensor
has the standard form
\begin{equation}
  T^{ab}(x) = m \int \dot{\gamma}^a \dot{\gamma}^b \delta(x,\gamma(t)) \rmd t,
\end{equation}
so that the mass and charge are concentrated on the same worldline.
It then follows from \eqref{LieGKF} that $T^{ab} \LieX g_{ab} =0$.
Applying \eqref{PBarAv},
\begin{equation}
  \rmd \ItP_\xi/\rmd s = q \xi^a \nabla_a (\Phi^{\mathrm{ext}}
  + \Phi_{\mathrm{R,DW}}^\rmSelf) |_{\gamma(s)} .
\end{equation}
Comparison with \eqref{pDotandPDot} shows that the torque vanishes.
The force is therefore
\begin{equation}
  F_a(s) = q \nabla_a (\Phi^{\mathrm{ext}}
  + \Phi_{\mathrm{R,DW}}^\rmSelf) |_{\gamma(s)} .
  \label{ForcePt}
\end{equation}
This does not completely determine the evolution of the particle's
linear momentum. It could still couple to the angular momentum.

Using \eqref{Torque}, one finds that
\begin{equation}
  \dot{S}_{ab} = 2 p_{[a} \dot{\gamma}_{b]} .
  \label{SDotPoint}
\end{equation}
A center-of-mass condition must be placed on $\Gamma$ in order to
solve this equation. It is possible to use \eqref{GammaBar} to
derive an expression for the difference $\dot{\gamma}^a - p^a/m$
assuming that $p^a \dot{\gamma}_a = -m$ \cite{Dix79,EhlRud}. Suppose
that the spin vanishes at least instantaneously. Given that $N_{ab}
=0$, one then finds that (unsurprisingly)
\begin{equation}
  p^a = m \dot{\gamma}^a   .
  \label{CMPoint}
\end{equation}
Substituting this into \eqref{SDotPoint} shows that $\dot{S}_{ab} =
0$. We have derived in a rather pedantic way the fact that the
angular momentum of a point particle vanishes for all time if it
does so at any instant.

Taking advantage of this, the motion of a nonspinning body is
completely determined by \eqref{ForcePt} and \eqref{CMPoint}. The
gradient of the regular self-field was derived in \cite{Quinn}.
Substituting appropriately,
\begin{eqnarray}
  \frac{D}{\rmd s} ( m \dot{\gamma}^a )  =& q \nabla^a \Phi_{\mathrm{ext}} + q^2 \Big[ \frac{1}{3} h^{a}{}_{b} \left( \dddot{\gamma}^a
  + \frac{1}{2} R_{bc} \dot{\gamma}^c \right) \nonumber \\
  & ~ - \frac{1}{12} R \dot{\gamma}^a + \lim_{\epsilon \rightarrow 0}
  \int_{-\infty}^{s-\epsilon} \nabla^a \Gret(\gamma, \gamma') \rmd s' \Big].
  \label{PointMotion}
\end{eqnarray}
Equations governing $\ddot{\gamma}^a$ and $\dot{m}$ are easily
extracted using the projection operator
\begin{equation}
  h^{a}{}_b = \delta^a_b + \dot{\gamma}^a \dot{\gamma}_b.
\end{equation}
This result is standard, and has been found in the past using
several different methods
\cite{DetWhiting,Quinn,RosenthalMassField}. The derivation here
starts from a formalism that's exact for any finite body. The only
external assumption required to include point particles was that the
momentum could be renormalized via \eqref{PRen}. The portion of the
singular self-field not taken into account with this procedure was
shown to be irrelevant to the body's motion. These results were
obtained without any detailed calculations of the singular
self-field. This is convenient, as its structure is usually much
more complicated than that of $\Phi^\rmSelf_{\mathrm{R}}$.

\subsection{A scaling limit} \label{Sect:Scaling Limit}

Given the definitions \eqref{PDefInt} and \eqref{EDef} for
$\ItP_\xi$ and $\mathcal{E}_\xi$, it is not completely consistent to
assume that the effective momenta $\hat{\ItP}_\xi$ remain finite for
distributional sources like \eqref{PointParticle}. The laws of
motion in Sect. \ref{Sect:ScalarForces} were derived under the
assumption that $\rho$ is well-behaved. It is not clear that they
can be used to discuss the behavior of singular charge
distributions. Furthermore, the direct use of point charges often
loses all sense of mathematical meaning when considering couplings
to fields that satisfy nonlinear wave equations. This problem is
particularly well-known when trying to discuss gravitational
self-forces \cite{GerochTraschen}.

Despite these remarks, point particles are introduced in practice
(with some special rules) in order to simplify calculations. They
are intended to represent the behavior of ``sufficiently small''
extended charges in an appropriate sense. Understanding this
equivalence and its limits is difficult, although it is relatively
straightforward to comment on \textit{a} particular class of
extended charges whose behavior approaches that of a point particle.
The techniques already developed in Sect. \ref{Sect:Point Particles}
generalize fairly easily to the discussion of a scaling limit.

Consider a one-parameter family of charge distributions
$\rho(x;\lambda)$ with diameters proportional to $\lambda$. These
objects shrink into $\Gamma$ as $\lambda \rightarrow 0$. The total
charge cannot remain fixed in this limit if $\mathcal{E}_\xi$ is to
remain well-defined. Suppose that
\begin{equation}
  \rho = \lambda^{-\alpha} \bar{\rho}(\mathbf{r}/\lambda, s_0 +
  s/\lambda) ,
\end{equation}
where $\mathbf{r}$ and $s$ represent Fermi normal coordinates
constructed using the center-of-mass as an origin. An appropriate
choice for the constant $\alpha$ is not obvious, so it will be left
free for now. Assume that the stress-energy tensor shrinks like the
charge density, but that its magnitude is made proportional to
$\lambda^{-\beta}$ rather than $\lambda^{-\alpha}$. Whatever $\beta$
happens to be, the bare mass scales like $\lambda^{3-\beta}$ to
leading order.

Each contribution to the laws of motion affecting this family of
charges scales differently as $\lambda \rightarrow 0$. If the
external scalar field remains finite in this limit, the force that
it exerts satisfies
\begin{equation}
  \int_\Sigma \rho \LieX \Phi^{\mathrm{ext}} t^a \rmd S_a \sim
  \Or(\lambda^{3-\alpha}) . \label{PhiExt}
\end{equation}
This is assumed to be the dominant influence on a body's motion when
it is sufficiently small. One additional power of $\lambda$ appears
if the GKF vanishes at the appropriate point on $\Gamma$. The torque
therefore scales like $\lambda^{4-\alpha}$.

The magnitude of the scalar self-force arising from the body's
regular self-field can be estimated from its point particle
expansion. This was used in \eqref{PointMotion}. If the result there
can be considered approximately valid inside a slowly-evolving
extended charge distribution (as has been verified directly in
electromagnetism \cite{HarteEM}), $\LieX
\Phi_{\mathrm{R,DW}}^\rmSelf$ depends on $q \dddot{\gamma}^a$ and
$q/\mathcal{R}^2$, where $\mathcal{R}$ is the curvature radius.
Given that the background geometry is assumed to be independent of
the charge's existence, $\mathcal{R}$ doesn't depend on $\lambda$.
The rate of change of acceleration can be estimated from
\eqref{PhiExt}. These two contributions sometimes scale differently,
so
\begin{equation}
  \int_\Sigma \rho \LieX \Phi_{\mathrm{R,DW}}^\rmSelf t^a \rmd S_a \sim
  \Or(\lambda^{6+\beta-3\alpha} + \lambda^{6-2\alpha}) .
  \label{SelfForceR}
\end{equation}
Which of these estimates dominates depends on whether or not
$\alpha$ is larger than $\beta$. The center-of-mass acceleration
remains finite as $\lambda \rightarrow 0$ when $\beta \geq \alpha$.
It is reasonable to suppose that this is always true, in which case
the self-force due to $\Phi^{\mathrm{R,DW}}_\rmSelf$ is always
proportional to $\lambda^{6-2 \alpha}$.

There is also a self-force due to the singular component of the
self-field. Slightly generalizing results used in the point particle
case,
\begin{equation}
  \LieX \sigma \sim \Or(\lambda^4) , \qquad \LieX \ln \Delta \sim
  \Or(\lambda^2) .
\end{equation}
It follows that
\begin{equation}
   \int_\Sigma \rmd S_a t^a \int \rmd V' \rho  \rho' \LieX G_{\mathrm{S,DW}}
   \sim \Or (\lambda^{7-2 \alpha}) . \label{SelfForceLie}
\end{equation}
This will always decrease faster than the regular component of the
self-force as $\lambda \rightarrow 0$. The Detweiler-Whiting axiom
is therefore satisfied for all sufficiently small and slowly varying
charge distributions.

This does not guarantee that $\Gamma$ will evolve like the worldline
of a point particle with the appropriate mass and charge. Strictly
adhering to the point particle equations of motion requires placing
several restrictions on the two scaling parameters $\alpha$ and
$\beta$. One of these comes from demanding that any deviations from
$\dot{p}^a \simeq \mathrm{D} (m \dot{\gamma}^a )/\rmd s$ be small
compared to the regular component of the self-force. Such terms can
scale like $\lambda^{4-\alpha}$, so let
\begin{equation}
  \alpha > 2. \label{AlphaRestrict}
\end{equation}
This same condition also arises if the interaction of the charge's
dipole moment with the external field is assumed to be negligible.

The magnitude of the gravitational force arising from $T^{ab} \LieX
g_{ab}$ must also be addressed. The gravitational force can be
estimated by
\begin{equation}
  \int_\Sigma T^{ab} \LieX g_{ab} t^a \rmd S_a \sim
  \Or(\lambda^{5-\beta}) . \label{GravForce}
\end{equation}
This follows from the fact that the $\LieX g_{ab}$ and its first
derivatives always vanish on the central worldline. It is reasonable
to suppose that the effective momenta are mainly determined by the
body's stress-energy tensor (rather than its self-field), so
$q^2/mD$ shouldn't increase as $\lambda \rightarrow 0$. This means
that
\begin{equation}
  \beta \geq 2 (\alpha -1). \label{BetaRestrict}
\end{equation}
If the mass is assumed to remain bounded -- meaning that $\beta \leq
3$ -- \eqref{AlphaRestrict} and \eqref{BetaRestrict} are more than
sufficient to guarantee that the gravitational force is negligible
compared to the self-force.

One last detail is the Papapetrou spin-curvature coupling in
\eqref{Force}. The body's angular momentum generically scales like
$\lambda^{4-\beta}$. The spin force shares this same behavior, and
can only be small compared to the regular self-force when $\beta < 2
(\alpha-1)$. This contradicts \eqref{BetaRestrict}. The
one-parameter families of charges considered here have the property
that either the Papapetrou force is important or the mass has very
little to do with $T^{ab}$. The latter possibility seems difficult
to accept, as any objects whose inertia was dominated by their
self-energy would probably be unstable or at least experience rapid
internal oscillations. It also isn't clear if the center-of-mass
conditions are meaningful in such cases. As a compromise, the
equality in \eqref{BetaRestrict} might be assumed to hold. This
means that $2< \alpha \leq 5/2$. It implies that fractional
self-energy remains finite as $\lambda \rightarrow 0$. Self-forces
have effects comparable to those of the angular momentum, so one
cannot be included without the other. Initial conditions might be
chosen such that the angular momentum can be ignored, although it is
not clear how long this condition could be kept consistent.

In conclusion, it is difficult to arrange all extended-body effects
to be negligible compared to those arising from the self-field. This
is especially true if the mass and charge densities are demanded not
to diverge as $\lambda \rightarrow 0$. Such a condition might be
required in order to maintain the test mass approximation that's
been assumed. Despite all of these remarks, it is rather trivial to
modify \eqref{PointMotion} to include charge dipole or mass
quadrupole effects. The main point of this discussion is really that
that the ``extended body self-force'' \eqref{SelfForceLie} does
generically become small compared to the regular self-force
\eqref{SelfForceR}. This is the main content of the
Detweiler-Whiting axiom applied to extended charges.


\section{Discussion}

Following \cite{HarteSyms,Dix74,Dix79}, approximate Killing fields
have been used to define the linear and angular momenta of extended
scalar charge distributions in curved spacetimes. These quantities
were shown to evolve according to \eqref{PBarAv}. The various terms
in that equation all have simple interpretations. A charge's
behavior is seen to have five distinct components. Two of these are
standard test body interactions with the background geometry and
external scalar field. The remaining three contributions to the
momentum evolution decompose the self-force and self-torque in a
particular way. The portion due to the regular self-field is
essentially as expected.

More interesting are the effects of the singular self-field. One
consequence of its presence is the introduction of what appear to be
effective linear and angular momenta connected to the scalars
$\mathcal{E}_\xi$. This is at least qualitatively an expected
result. Effective masses are usually found when introducing specific
charge distributions and using perturbation theory to approximate
their self-fields \cite{OriExtend,Thirring}. By contrast, the
definition \eqref{EDef} for the self-momentum obtained here required
only straightforward manipulations in the full theory. It includes a
number of effects more complicated than simple mass shifts. These
would probably not be obvious from an inspection of approximate
forces and torques. A related issue is that the effective momenta
introduce a possible ambiguity in determining a charge's motion.
Centers-of-mass might be defined using only the bare momenta defined
in terms of the stress-energy tensor, or using the full momenta
$\hat{\ItP}_\xi = \ItP_\xi + \mathcal{E}_\xi$. These two
possibilities generically lead to different worldlines. It is not
clear which -- if either -- is more appropriate for charges with
very large self-fields. The laws of motion simplify when the full
momenta are used to define a body's mass center. Many of the
unexpected results in \cite{HarteEM} regarding the behavior of
electromagnetic charges in flat spacetime can probably be attributed
to failing to fully apply this simplification.

The singular self-self field affects the force and torque more
directly as well. This arises from a term in \eqref{PBarAv}
involving $\LieX \GS$. It has the physical interpretation of
measuring the degree to which the singular self-field fails to
satisfy Newton's third law in the direction defined by $\xi^a$. It
is also related to the failure of this field to be conservative.
Comparison with \eqref{PDotSelfNewt} shows that such effects would
exist even if the field equation were elliptic. It therefore should
not be thought of as a reaction to emitted radiation. In general,
the Lie derivative of a singular Green function always satisfies
\begin{eqnarray}
  \Box \LieX \GS(x,x') = \big[ \nabla^a \nabla^b \GS + 2\pi
  \delta(x,x') g^{ab} \big] \nonumber
  \\ ~ \times \LieX g_{ab} + \big[ \nabla^b \LieX g_{ab} - \frac{1}{2} \nabla_a ( g^{bc} \LieX g_{bc} )
  \big] \nabla^a \GS.
  \label{BoxLieG}
\end{eqnarray}
The degree to which $\LieX g_{ab}$ remains small determines how
large the source terms on the right-hand side can be. By
construction, they always decrease near the worldline used to
construct the GKFs.

The results derived here provide a simple framework within which to
generalize the Detweiler-Whiting axiom for extended charge
distributions. Let this mean that the singular component of a body's
self-field -- as defined by the Green function \eqref{GSingularDW}
-- has no explicit effect on the evolution of the full momenta
$\hat{\ItP}_\xi$. It is equivalent to demanding \eqref{PointResult}
or an approximate equivalent. All real Killing fields are also GKFs,
so this result is exact for any charge distribution in the maximally
symmetric Minkowski or de Sitter spacetimes. If only one or a few
Killing fields exist, it is also exact for linear combinations of
the momenta with the form $\hat{p}^a K_a + \hat{S}^{ab} \nabla_{[a}
K_{b]}/2$. The results of Sect. \ref{Sect:Apps} extend the
Detweiler-Whiting axiom to be approximately valid in all spacetimes
when a charge's diameter is much smaller than the local curvature
scales. Extensions of these ideas to electromagnetic and
gravitational self-forces will be explored in future papers.

\ack

I am grateful for many helpful comments and discussions with Robert
Wald and Samuel Gralla. This work was supported by NSF grant
PHY04-56619 to the University of Chicago.

\end{document}